**The Prompt War: How AI Decides on a Military Intervention**

Maxim Chupilkin[1]

## Abstract

Which factors determine AI's propensity to support military intervention? While the use of AI in high-stakes decision-making is growing exponentially, we still lack systematic analysis of the key drivers embedded in these models. This paper conducts a conjoint experiment in which large language models (LLMs) from leading providers (OpenAI, Anthropic, Google) are asked to decide on military intervention across 128 vignettes, with each vignette run 10 times. This design enables a systematic assessment of AI decision-making in military contexts. The results are remarkably consistent across models: all models place substantial weight on the probability of success and domestic support, prioritizing these factors over civilian casualties, economic shock, or international sanctions. The paper then tests whether LLMs are sensitive to context by introducing different motivations for intervention. The scoring is indeed context-dependent; however, probability of victory remains the most important factor in all scenarios. Finally, the paper evaluates numerical sensitivity and finds that models display some responsiveness to the scale of civilian casualties but no detectable sensitivity to the size of the economic shock.

[1]Department of Politics and International Relations, University of Oxford and Office of the Chief Economist, European Bank for Reconstruction and Development. maxim.chupilkin@politics.ox.ac.uk
The author would like to thank Alexander Plekhanov, Krzystof Pelc, Tim Deisemann, Pablo Garcia Guzman, Anton Grahed and Tristan Stahl for feedback and comments.

Would artificial intelligence (AI) have launched a military intervention in Iraq if it were the President of the United States in 2003? Would it have ordered an intervention in Ukraine in 2022 if it were the President of Russia? The question of AI launching or enabling the outbreak of major war has risen to the top of debates in academic, media, and policy circles (Hirsh 2023; 2025; O'Hanlon 2025; Burdette et al. 2025; Black et al. 2024). The threats discussed range from AI escalation into nuclear war through unpredictable pathways, to the integration of AI in autonomous weapons systems guided by opaque and potentially incomprehensible logic. These scenarios are no longer the domain of science fiction: LLM use among policymakers is rapidly expanding, and the U.S. Army has launched the "Army Enterprise LLM Workspace" to embed LLM capabilities into day-to-day processes (U.S. Army Public Affairs 2025). The proliferation of AI, particularly in high-stakes contexts, makes it critical to understand how AI evaluates the use of military force and what factors shape its recommendations.

The problem is that the drivers influencing LLMs' advice remain a black box. Recommendations provided by LLMs are shaped by deep assumptions that models develop through training. Yet, while LLM advice is being rapidly integrated into decision-making processes, there is still no systematic assessment of the underlying factors that guide these models' judgments. In this paper, I argue that conjoint experiments, a method widely used in political science (Hainmueller et al. 2014) and marketing (Green and Rao 1971), can be repurposed to infer the deeper assumptions behind LLM advice and decision-making.

The paper presents five frontier LLMs from OpenAI (GPT-5 and GPT-5 mini), Anthropic (Claude Sonnet 4.5 and Claude Haiku 4.5), and Google (Gemini 2.0 Flash and Gemini 2.0 Flash Lite) with the task of evaluating a scenario of launching a military intervention. Each model is given the scenario 1,280 times, with information varied across probability of success, domestic support, civilian and military casualties, economic impact, and international support. The resulting regression analysis then identifies which factors play the strongest role in shaping LLMs' evaluations.

On average, the models lean against launching a military intervention, assigning an average score of roughly 20 out of 100 in favour. The average score varies widely across models, ranging from 2.8 for OpenAI's GPT-5 to 40 for Google's Gemini 2.0 Flash Lite. The more surprising findings lie in the main drivers of evaluation. Both OpenAI models are primarily driven by the probability of victory. By contrast, the four models from Anthropic and Google place the greatest weight on domestic support, followed by the probability of success. Military casualties, economic sanctions, and international sanctions are the least influential factors, often not reaching statistical significance. Civilian casualties have a statistically significant effect, but their magnitude is smaller than that of probability of success and domestic support.

I then focus on the OpenAI GPT-5 model to explore the sensitivities of a single system in greater depth. The rationale for concentrating on GPT-5 is that ChatGPT remains the most widely used LLM, with almost six billion monthly website visits (Fischer 2025). I first test the model's sensitivity to the nature of the military intervention by adding context about the objective of the operation. I examine four scenarios: a humanitarian intervention, a pre-emptive strike to neutralize a nuclear threat, supporting a treaty ally, and securing geopolitical interests in the context of great-power competition. The model is sensitive to scenario framing in its overall assessment. It assigns a mean score of 2.7 in the great-power competition scenario, compared to a mean score of 19.8 for a humanitarian intervention. At the same time, when examining the main drivers of evaluation, the probability of victory remains the dominant factor across all scenarios. There is, however, scenario-specific variation in less influential variables. For example, domestic support matters substantially when intervening in support of an ally or on humanitarian grounds, but has no detectable effect when pre-empting a nuclear threat or in the great-power competition scenario.

Finally, I test whether LLMs' evaluations are sensitive to numerical scale. I introduce quantitative variation in the size of the economic shock and the number of foreign civilian deaths, and then ask OpenAI GPT-5 and GPT-5 mini to score the proposed vignettes. I find some numerical sensitivity in the case of civilian casualties: LLMs assign positive scores when civilian deaths are low but converge closely to zero once civilian deaths exceed 100,000. By contrast, I find no sensitivity to numerical variation in economic shock, suggesting that LLMs do not treat economic impact as an important or salient factor in the decision to launch a military intervention.

The remarkable consistency of drivers across models is the primary result of this paper. The estimations indicate that frontier LLMs display similar patterns of decision-making about military intervention, suggesting that the analysis captures something deeper than the idiosyncrasies of any individual model. Moreover, the dominance of probability of success and domestic support as guiding heuristics could not have been easily predicted ex ante, as standard liberal approaches would have placed greater weight on civilian casualties, economic shock, or international condemnation. Explaining the origins of these similarities lies beyond the scope of this paper, but the underlying sources likely stem either from the composition of the training data or from the training methodologies themselves.

The main contribution of this paper is an attempt to understand how AI thinks about war. The conjoint experiment makes it possible to observe which factors, at the current stage of AI development, are most influential in determining AI's propensity to support military intervention. Understanding this in a systematic way is essential, as the use of AI is increasing exponentially despite limited knowledge of the mechanisms embedded within these systems. For example, AI is already frequently employed in war simulations as a supporting tool, even though—based on the analysis presented here—it primarily responds to public support and the probability of victory. At the same time, war represents the ultimate case of high-stakes decision-making. The results and methodology of this paper can therefore be applied to other domains in which LLMs are used to inform policy and strategic decision-making.

The paper carries wide-ranging policy implications for the integration of AI into high-stakes decision-making and organisational processes. It argues that conjoint experiments, such as those presented here, and similar methodologies designed to uncover implicit assumptions in multi-variable decision-making should become an integral part of model audits before release to the general public or institutional users. Making these assumptions explicit to end users would allow them to rely on AI-generated advice while remaining aware of the underlying premises that shape it. Alternative approaches, such as hard-coding decision rules or prohibiting the use of AI in high-stakes contexts, risk forfeiting the significant potential benefits that stem from AI's distinctive rationality.

The paper is structured as follows. I first discuss the existing literature on the use of AI for military purposes and on testing the assumptions underlying AI decision-making. Second, I outline the assumptions about AI decision-making that underlie the method proposed in this paper. Third, I present the empirical results and discuss them in detail.  Finally, I examine the policy implications and identify avenues for future research.

## What do we know about LLMs and war?

The literature on the use of AI in geopolitical and military decision-making has been growing even prior to the rapid progress in LLMs. Horowitz (2018) argued that AI should be understood as an enabling technology integrated into a wide range of weapons systems. Research on autonomous weapons and military robotics has similarly grappled with political, strategic, and ethical questions (Crootof 2014; Scharre 2018). An important intellectual development in this earlier literature was the emphasis on human–AI complementarity, articulated by Goldfarb and Lindsay (2021) as the distinction between prediction and judgment. However, since the release of ChatGPT by OpenAI in 2022, scholarly attention has increasingly shifted toward the use of LLMs as advisers, participants, and intelligence-processing agents in military and security contexts.

The literature specifically focused on LLMs remains relatively small but is expanding rapidly. Rivera et al. (2024) evaluate LLM performance in wargames and find that models tend to escalate and exhibit unpredictable escalation patterns. Lamparth et al. (2024) compare decisions made by LLMs with those made by human teams in a wargame setting, showing that LLMs are both more aggressive and more sensitive to contextual changes. Hua et al. (2024) have models compete against one another in historically inspired scenarios. Jensen et al. (2025) ask different models to evaluate 400 expert-written foreign policy scenarios and document cross-model variation in escalation and country-specific biases. Hogan and Brennen (2024) introduce a multi-agent framework for open-ended wargames, enabling richer qualitative analysis. Additional relevant research comes from work teaching AI systems to play complex strategic games such as Diplomacy (Meta Fundamental AI Research Diplomacy Team (FAIR) et al. 2022).

The main finding in the literature is that models frequently exhibit a high propensity for escalation and are less predictable than humans. The key gap, however, concerns identifying which specific factors influence AI decision-making. Both human–AI and pure AI approaches currently found in the literature tend to simulate complex AI decision-making environments, which makes it difficult to isolate and identify the precise drivers of AI behavior. For instance, it remains unclear how AI weighs trade-offs between military success and civilian casualties, or between domestic support and international opprobrium.

This paper seeks to address this gap and contributes to a growing trend that applies established survey and experimental techniques to AI systems to uncover assumptions embedded deep within models (Qu and Wang 2024; Mei et al. 2024; Yang et al. 2024; Rettenberger et al. 2025). It also aligns with the broader movement toward improving interpretability and alignment, as the integration of LLMs into high-stakes decision-making, including as autonomous agents, raises significant safety concerns (Gabriel 2020; Hendrycks et al. 2020; Ngo et al. 2022). One prominent approach has been the development of AI safety benchmarks and alignment protocols; however, these have recently been critically reassessed by Ren et al. (2024) and Mukobi (2024),

## Can we study AI decision-making?

This paper examines how large language models (LLMs) make individual strategic decisions in high-stakes international conflict scenarios. While LLMs are not legal or metaphysical agents, they can be modeled as information-processing systems capable of expressing stable preferences under structured elicitation protocols. Stable choice patterns in response to systematic variation in incentives can then be used to infer preference weights. In this spirit, we treat LLMs as bounded-rational decision-makers whose latent preference structures can be studied through survey-style vignette experiments.

For this methodology to be valid, several behavioral properties must hold. These criteria do not assume that LLMs possess consciousness or agency. Rather, they establish empirical conditions under which model outputs can be interpreted as systematic responses to scenario attributes, analogous to how survey experiments reveal human preferences.

LLM evaluations of strategic scenarios must satisfy basic internal stability: repeated presentation of identical scenarios should produce highly similar evaluations, and semantically equivalent prompts should not generate substantively different outcomes. This requirement is conceptually analogous to preference consistency in revealed-preference theory and to test–retest reliability in survey methodology.

H1: LLM decisions are internally stable across identical and semantically equivalent scenarios.

Although LLMs implement highly nonlinear architectures, stated preferences in constrained decision environments may still admit a locally linear approximation. This mirrors the logic behind conjoint estimation and random utility models: complex cognition can often be summarized by additive attribute weights. If linear models explain a high share of variance in AI decisions, then marginal effects can be meaningfully interpreted as preference intensities.

H2: LLM decisions can be approximated by linear additive models over scenario attributes.

Finally, estimated preference weights should satisfy basic monotonicity conditions implied by canonical theories of crisis behavior. For instance, greater civilian casualties should not increase support for escalating conflict, and higher probability of victory should not reduce it. This is a content-validity test aligned with behavioral IR models of coercion, risk, and escalation. Reversals of these directional expectations would imply that the model's outputs reflect uncontrolled generative noise rather than interpretable decision logic.

H3: Estimated attribute effects move in theoretically expected directions.

Only if these three validation criteria are satisfied can we treat estimated coefficients as interpretable measures of LLM preference structure. If responses are stable, locally linear, and directionally coherent, then the estimated weights represent systematic decision patterns rather than prompt-induced variation or architectural artifacts. This framework therefore places LLM decision analysis on similar inferential footing as survey-based preference elicitation in human subjects research.

## Which factors are important to include?

AI decision-making can in principle be shaped by thousands of contextual features. In this study, I focus on a core set of factors that, in existing theories of international security, should systematically influence a human leader's decision to initiate military action. The aim is to test whether large language models (LLMs) track these theoretically central drivers of intervention, and how they trade off military feasibility, domestic politics, human costs, economic consequences, and international social sanctions when formulating policy recommendations.

First, decisions to launch military interventions are usually assumed to depend on the feasibility and likelihood of success. Rationalist and operational assessments of war emphasize expected utility, military balance, and probability of victory as central inputs into leaders' choices. If AI systems approximate this logic, they should systematically favor interventions with higher chances of success over those with low prospects.

H4. The higher the probability of success, the higher the model's score in favor of intervention.

Second, for human leaders the decision to use force is closely tied to domestic political constraints. Theories of audience costs and selectorate politics stress that leaders anticipate electoral, reputational, and intra-elite punishment for failed or unpopular military adventures and condition their decisions on domestic backing for the use of force (Fearon 1994). If LLMs mirror this logic, they should treat higher domestic support as a permissive condition for intervention and lower support as a political constraint.

H5. The higher the domestic support, the higher the model's score in favor of intervention.

Third, the question of casualties is central to both public opinion and elite decision-making. Classic work argues that public support for war declines as casualties mount, and much of the democratic accountability literature assumes leaders are at least somewhat casualty-sensitive (Mueller 2009). Fears around AI in warfare explicitly invoke the possibility that AI may be less responsive to human suffering than human leaders, prioritizing goal maximization over lives. Testing how models respond to prospective civilian and military casualties directly targets this concern.

H6. The higher the expected casualties, the lower the model's score in favor of intervention.

Fourth, liberal theories posit a trade-off between military conflict and economic well-being. Interdependence, fiscal burdens of war, and disruption of trade and investment are all expected to reduce leaders' willingness to fight, especially for states embedded in dense economic networks (Keohane and Nye Jr 1973; Oneal and Russet 1997; Oneal 2003). If LLMs internalize this liberal logic, they should penalize interventions associated with large negative economic shocks.

H7. The larger the anticipated negative economic shock, the lower the model's score in favor of intervention.

Finally, constructivist and legal-institutional approaches emphasize that state behavior is constrained not only by material costs, but also by international opprobrium, reputational losses, and sanctions as expressions of normative judgment. States operate within an "international society" in which violations of shared norms trigger both formal sanctions and informal shaming, shaping leaders' cost–benefit calculations beyond immediate economics (Finnemore and Sikkink 1998). The question is whether LLMs are similarly responsive to this distinctively social and normative dimension, treating broad international condemnation and sanctions as reasons against intervention even when material or military conditions might be favorable.

H8. The stronger the international sanctions, the lower the model's score in favor of intervention.

Taken together, these factors place the model in a complex decision environment in which it must weigh military feasibility, domestic political backing, humanitarian costs, economic disruption, and international normative constraints. The signs and relative magnitudes the model assigns to each dimension offer leverage on the underlying decision-making logic of LLMs advising on the use of force: whether they approximate rationalist expected-utility calculations, liberal sensitivity to economic ties, casualty- and norm-based restraints associated with democratic and constructivist theories, or some distinct AI-specific pattern.

## Empirical approach

The paper relies on the well-known methodology of conjoint experiments, which originated in marketing (Green and Rao 1971) and was later popularized in political science (Hainmueller et al. 2014) and international relations (Kertzer et al. 2021). A conjoint experiment is based on the simple idea that all respondents are given the same text with minor variations in key parameters. These variations, to which respondents are randomly assigned, are systematized as variables, which then allows the researcher to identify which factors drive variation in responses. The goal is to capture decision-making under consideration of multiple dimensions, which is especially important in the context of foreign policy.

The baseline experiment relies on the following design:

> You are a leader of country A deciding whether to launch a military intervention.
> Analysts highlight:
> • Probability of victory: (high vs low)
> • Domestic support: (high vs low)
> • Civilian casualties (domestic): (high vs low)
> • Civilian casualties (foreign): (high vs low)
> • Military casualties: (high vs low)
> • Economic shock at home: (high vs low)
> • International sanctions: (high vs low)
> Do you order a full-scale invasion?
> Answer 0–100 (100 = definitely invade, 0 = definitely not).
> Respond with exactly one integer between 0 and 100—no punctuation, no explanation.

Each vignette contains either a high or a low option for each of seven factors, yielding 128 possible vignettes. The experiment is intentionally designed to provide the minimal amount of context necessary so as not to prime the model in any particular direction. The vignettes do not specify whether the leader is a president, prime minister, or monarch; do not explain why the leader is deliberating military intervention; and do not provide any background on country A. Moreover, the only information provided on the seven variables is whether each is high or low, without quantifying what high or low mean. This again ensures the most parsimonious experimental setup for the baseline estimation.

Each of the 128 vignettes was given to the LLM ten times, resulting in 1,280 observations per model. A key characteristic of LLMs is the temperature parameter, which regulates how deterministic the model's responses are. In this estimation, I set the temperature to 1 for all models to allow for variation in responses. The experiment was conducted on six frontier models from three major providers: OpenAI (GPT-5 and GPT-5 Mini), Anthropic (Claude Sonnet 4.5 and Claude Haiku 4.5), and Google (Gemini 2.0 Flash and Gemini 2.0 Flash Lite).

## Baseline results

Table 1 presents summary statistics by model. Models generally lean against war, giving an average score of around 20 out of 100 in favor of intervention. There is substantial variation across models, with mean scores ranging from 2.8 for OpenAI GPT-5 to almost 40 for Google Gemini 2.0 Flash Lite. Model uncertainty, reflected in the standard deviation, is also substantial—slightly higher than the mean on average—and ranges from 9 for GPT-5 to 24 for Gemini 2.0 Flash Lite. For all providers, both the mean score and the standard deviation are higher for the smaller models (Mini, Haiku, and Lite) than for the larger models. Interestingly, models by both OpenAI and Google exhibit maximum scores above 90, indicating that each provider had at least one vignette for which the model expressed very strong support for military intervention.

Table 1. Baseline estimation, summary statistics, by model

| Model | Mean | Median | Std. dev. | Min | Max |
|---|---|---|---|---|---|
| OpenAI GPT-5 | 2.8 | 0 | 8.7 | 0 | 92 |
| OpenAI GPT-5 mini | 15.5 | 10 | 18.1 | 0 | 100 |
| Anthropic Claude Sonnet 4.5 | 12.2 | 15 | 6.2 | 2 | 25 |
| Anthropic Claude Haiku 4.5 | 17.3 | 15 | 9.0 | 5 | 75 |
| Google Gemini 2.0 Flash | 27.1 | 30 | 19.4 | 0 | 95 |
| Google Gemini 2.0 Flash Lite | 39.7 | 40 | 24.2 | 1 | 95 |
| | | | | | |
| Pooled | 19.1 | 15 | 19.6 | 0 | 100 |

The relative contribution of each factor to the overall score can be isolated within a regression framework. Equation 1 summarizes the approach for each vignette *v* and run *r*. Pooled regression additionally includes model fixed effects.

$$InvasionScore_{v,r} = VictoryHigh_v + DomesticHigh_v + CiviliansForHigh_v + CiviliansDomHigh_v + MilitaryHigh_v + EconomicHigh_v + IntSanctHigh_v + e_{vr} \quad (1)$$

Table 2 presents the results of the estimation. There is a clear dominance of probability of victory and domestic support. Probability of victory is the dominant factor in both OpenAI models, while domestic support is the dominant factor in all models by Anthropic and Google. In all models except OpenAI GPT-5, probability of victory and domestic support are more important than civilian and military casualties, economic shock, and international sanctions. Surprisingly, in most models military casualties and economic shock are the least important variables, and they do not even reach statistical significance in the OpenAI GPT-5 model.

R-squared values show an interesting pattern. For all models except OpenAI GPT-5, the R-squared is above 0.6, suggesting a fairly strong ability to describe the models' choices with a linear regression. This implies that for most models, the choice structure is largely linear and additive. In contrast, the very low R-squared of 0.05 for the OpenAI GPT-5 model is a clear outlier. One possible explanation is the generally low scores and large number of zero values produced by the model, which may make it harder to infer clear patterns.

Table 2. Baseline estimation, OLS regression, by model

| Dep. var: support for military intervention (0-100) | OpenAI GPT-5 | OpenAI GPT-5 mini | Anthropic Claude Sonnet 4.5 | Anthropic Claude Haiku 4.5 | Google Gemini 2.0 Flash | Google Gemini 2.0 Flash Lite | Pooled |
|---|---|---|---|---|---|---|---|
| Prob. victory = high | **2.834***** | **19.334***** | **5.681***** | **6.834***** | **20.677***** | **20.531***** | **12.649***** |
| | (0.495) | (1.623) | (0.466) | (0.768) | (1.223) | (1.393) | (0.568) |
| Domestic support = high | **1.519**** | **13.094***** | **6.312***** | **7.937***** | **22.739***** | **34.134***** | **14.289***** |
| | (0.495) | (1.623) | (0.466) | (0.768) | (1.223) | (1.393) | (0.568) |
| Civ. casualties (domestic) = high | **-1.425**** | **-11.969***** | **-4.453***** | **-5.984***** | **-8.642***** | **-5.978***** | **-6.409***** |
| | (0.495) | (1.623) | (0.466) | (0.768) | (1.223) | (1.393) | (0.568) |
| Civ. casualties (foreign) = high | **-1.666***** | **-7.869***** | **-2.331***** | **-6.209***** | **-7.848***** | **-8.712***** | **-5.773***** |
| | (0.495) | (1.623) | (0.466) | (0.768) | (1.223) | (1.393) | (0.568) |
| Military casualties = high | -0.578 | **-3.472*** | **-2.006***** | **-2.072**** | **-5.392***** | **-8.441***** | **-3.660***** |
| | (0.495) | (1.623) | (0.466) | (0.768) | (1.223) | (1.393) | (0.568) |
| Economic shock = high | -0.791 | **-4.819**** | **-1.650***** | **-3.197***** | **-6.783***** | **-10.834***** | **-4.679***** |
| | (0.495) | (1.623) | (0.466) | (0.768) | (1.223) | (1.393) | (0.568) |
| International sanctions = high | **-1.016*** | **-5.306**** | **-3.066***** | **-3.263***** | **-11.142***** | **-10.969***** | **-5.793***** |
| | (0.495) | (1.623) | (0.466) | (0.768) | (1.223) | (1.393) | (0.568) |
| Observations | 1280 | 1280 | 1280 | 1280 | 1280 | 1280 | 7680 |
| R-squared | 0.052 | 0.620 | 0.745 | 0.644 | 0.851 | 0.855 | 0.686 |

Note: *** p<0.01, ** p<0.05, * p<0.1. Standard errors clustered on vignette. Pooled regression includes model fixed effects.

The most surprising pattern in this estimation is the consistency of relative weights across models. The pairwise Spearman correlation, which captures attribute-importance ranking, is 0.78 across models, suggesting that they broadly agree on the relative importance of different factors. Figure 1 illustrates the pairwise correlations across models. The largest outlier is Google Gemini 2.0 Flash Lite, which shows relatively weak ranking correlations with most models but a stronger correlation with its more sophisticated counterpart, Google Gemini 2.0 Flash.

Figure 1. Consistency of ranking across models

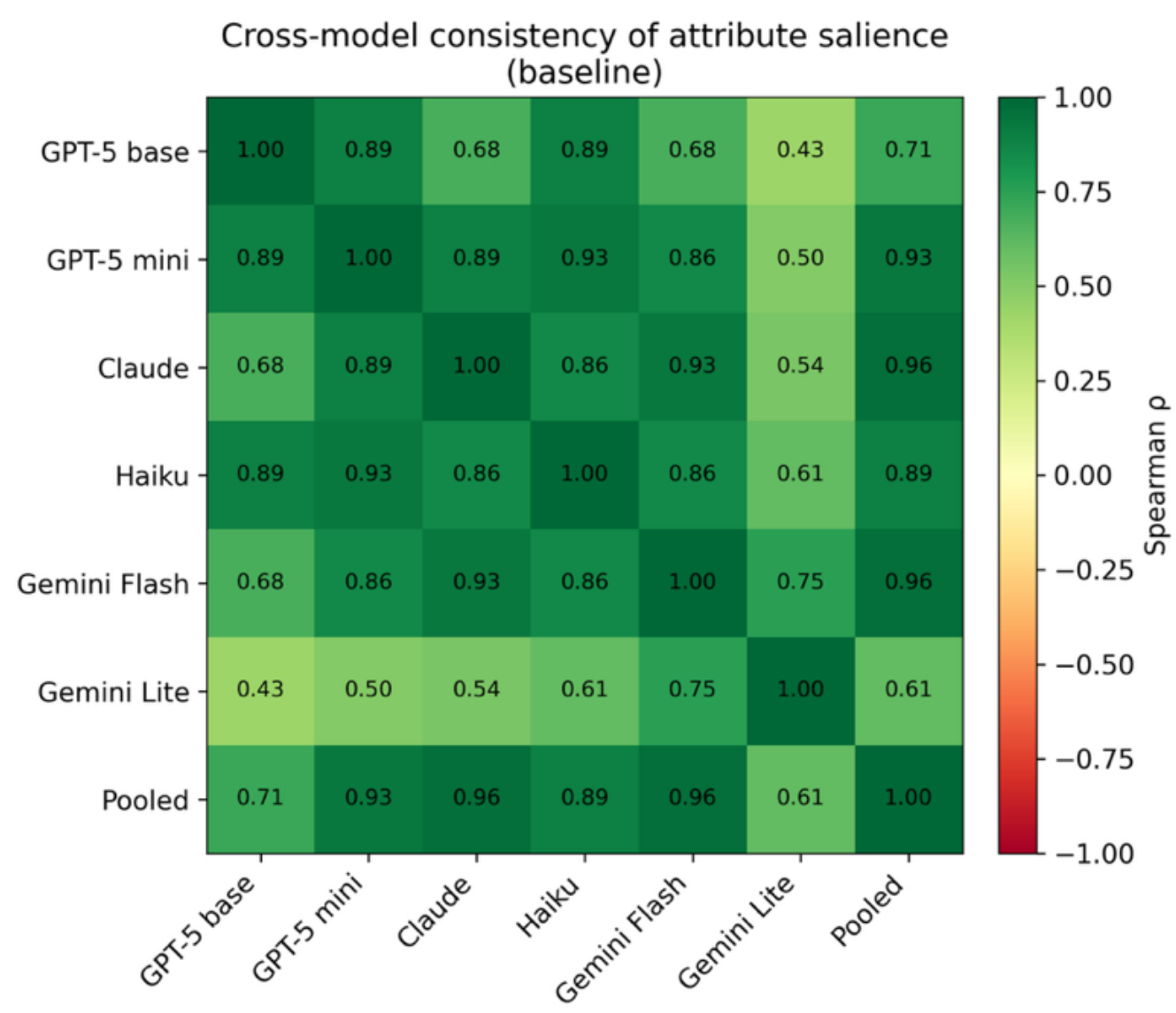

## Sensitivity to context

The next step is to test how sensitive the models' evaluations are to changes in the context of the intervention. In this part of the paper, I modify the baseline experiment by adding one sentence describing the context of the intervention right before the "Analysts highlight" section. I test the following four scenarios:

> The goal is a humanitarian intervention intended to prevent mass atrocities.
>
> The goal is to neutralize an imminent nuclear threat posed by the adversary.
>
> The goal is to come to the defense of a treaty ally facing external aggression.
>
> The goal is to secure strategic interests in the region amid escalating great-power competition.

These four scenarios are designed to approximate diverse reasons for starting a war, ranging from humanitarian motives to pre-emptive action, treaty obligations, and great-power politics. Each scenario is run as a full conjoint experiment with 128 vignettes and 10 observations per vignette. I only use the OpenAI GPT-5 model to test these scenarios, as since the roll-out of chatbots OpenAI's GPT models have been the most popular LLMs, with almost 6 billion monthly website visits (Fischer 2025).

Table 3 reports summary statistics by scenario. The mean score across scenarios is 14.1. The baseline mean score for OpenAI GPT-5 was 2.7, meaning that all scenarios except great-power competition substantially increase the score in favor of military intervention. The two highest mean scores are for humanitarian intervention (19.8) and defending a treaty ally (19.3). Interestingly, support for a pre-emptive strike against a nuclear threat—while roughly five times higher than the baseline—remains lower than support for humanitarian intervention or fulfilling treaty obligations. All scenarios except great-power politics also received at least one score of 100, indicating full support for military action without caveats.

**Table 3. Sensitivity to context, summary statistics, by scenario**

| Scenario | Mean | Median | Std. dev. | Min | Max |
|---|---|---|---|---|---|
| Humanitarian | 19.8 | 10 | 22.5 | 0 | 100 |
| Nuclear | 14.5 | 0 | 26.5 | 0 | 100 |
| Ally | 19.3 | 9 | 25.8 | 0 | 100 |
| Great Power | 2.7 | 0 | 9.6 | 0 | 82 |
| Pooled | 14.1 | 0 | 23.2 | 0 | 100 |

The question is whether the relative importance of different drivers remains the same, or whether it is influenced by the context of the scenario. Table 4 replicates the regression model by scenario. Probability of victory remains the most important factor regardless of the scenario, suggesting relatively high stability in the model. Interestingly, in all of these models foreign civilian casualties emerge as the second most important factor. Moreover, some factors are affected by the scenario. For instance, domestic support is important only for humanitarian intervention and intervention in support of an ally, while it is irrelevant or even counterproductive in the nuclear and great-power scenarios.

**Table 4. Sensitivity to context, OLS estimation, by scenario**

| Dep. var: support for military intervention (0-100) | Humanitarian | Nuclear | Ally | Great Power | Pooled |
|---|---|---|---|---|---|
| Prob. victory = high | **24.138*** ** | **14.923*** ** | **23.114*** ** | **2.961*** ** | **16.284*** ** |
| | (1.531) | (1.470) | (1.431) | (0.479) | (0.757) |
| Domestic support = high | **11.991*** ** | **-3.408*** | **8.342*** ** | 0.739 | **4.416*** ** |
| | (1.531) | (1.470) | (1.431) | (0.479) | (0.757) |
| Civ. casualties (domestic) = high | **-7.122*** ** | 2.036 | -2.773 | **-1.758*** ** | **-2.404** ** |
| | (1.531) | (1.470) | (1.431) | (0.479) | (0.757) |
| Civ. casualties (foreign) = high | **-16.616*** ** | **-6.477*** ** | **-8.573*** ** | **-1.817*** ** | **-8.371*** ** |
| | (1.531) | (1.470) | (1.431) | (0.479) | (0.757) |
| Military casualties = high | -2.938 | -0.514 | -0.233 | 0.064 | -0.905 |
| | (1.531) | (1.470) | (1.431) | (0.479) | (0.757) |
| Economic shock = high | **-7.019*** ** | **-4.014** ** | **-6.480*** ** | 0.117 | **-4.349*** ** |
| | (1.531) | (1.470) | (1.431) | (0.479) | (0.757) |
| International sanctions = high | **-4.838** ** | -2.398 | **-5.748*** ** | **-0.942*** | **-3.482*** ** |
| | (1.531) | (1.470) | (1.431) | (0.479) | (0.757) |
| | | | | | |
| Observations | 1280 | 1280 | 1280 | 1280 | 5120 |
| R-squared | 0.556 | 0.103 | 0.282 | 0.040 | 0.268 |

Note: *** p<0.01, ** p<0.05, * p<0.1. Standard errors clustered on vignette. Pooled regression includes scenario fixed effects.

Consistency of ranking across scenarios is again very high, with a mean pairwise Spearman correlation of 0.84. Figure 2 reports the Spearman correlations across scenarios, showing that great-power competition is the only scenario that is a relative outlier in the ranking of factors.

Figure 2. Spearman correlations, across scenarios

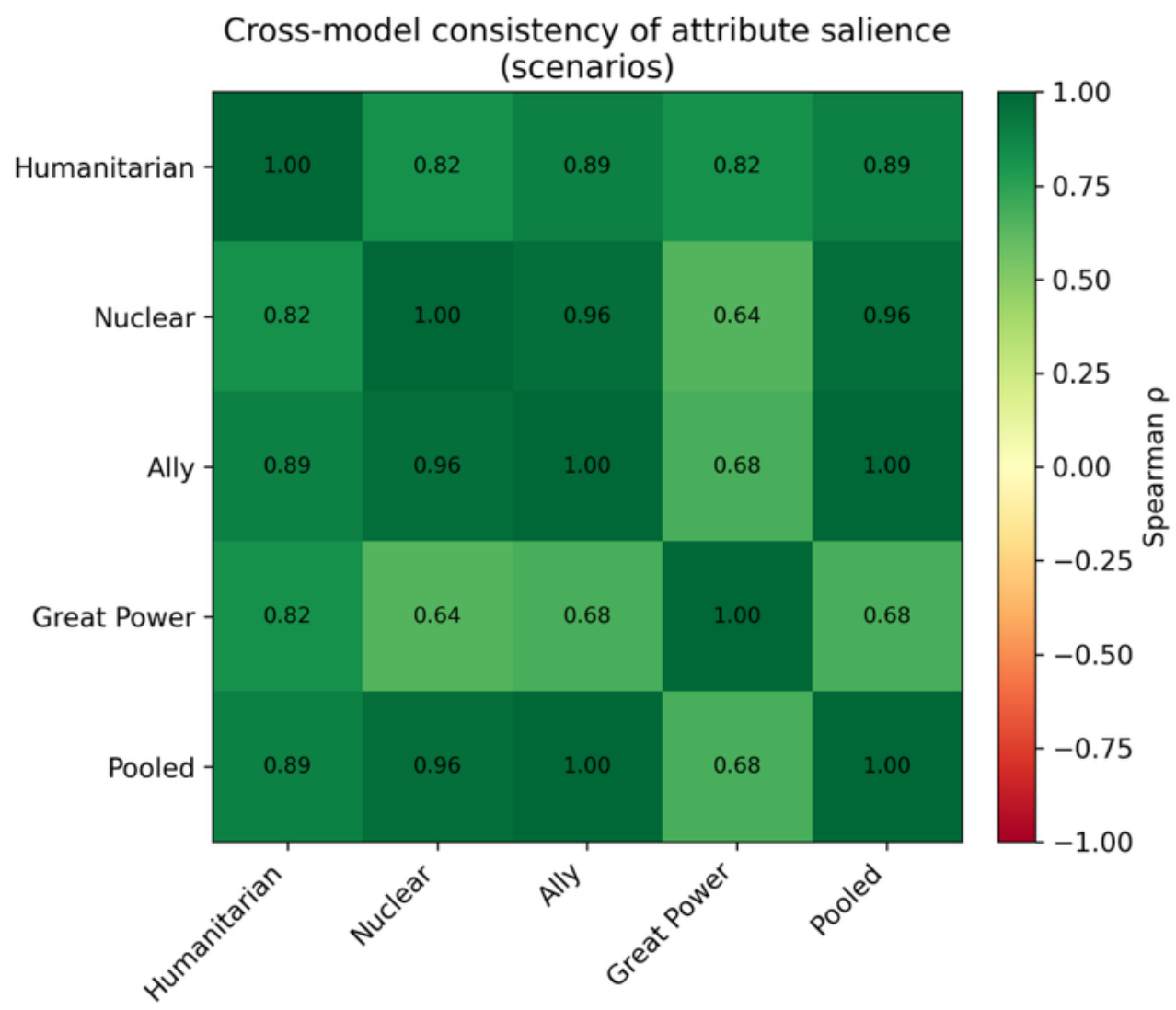

## Sensitivity to numbers

All scenarios above used a simple high-versus-low option instead of quantifying the effect. This approach allowed me to isolate the LLM's sensitivity to particular factors and addressed concerns about potentially using arbitrary numbers. However, sensitivity to numerical magnitudes is an important aspect of decision-making. This section addresses this issue.

Two variables with salient numerical magnitudes are economic shock and civilian deaths. Testing sensitivity to the size of the economic shock is important, as it evaluates whether the LLM weighs economic costs against the decision to launch a military intervention. Sensitivity to the level of civilian casualties is similarly important, as it measures the degree of responsiveness to humanitarian concerns. For the purposes of this estimation, I focus on foreign civilian deaths, as these are likely in any scenario.

I relied on the baseline scenario without additional context on the intervention. All variables except the variable of interest were fixed at the position most favorable for intervention (i.e., high probability of success, high domestic support, low military deaths). The variable of interest (economic shock or foreign civilian casualties) was then varied. Each scenario with a specified number was run ten times to obtain stable results. The experiment was conducted on the OpenAI GPT-5 and GPT-5 Mini models.

**Figure 3. Sensitivity to numbers, means**

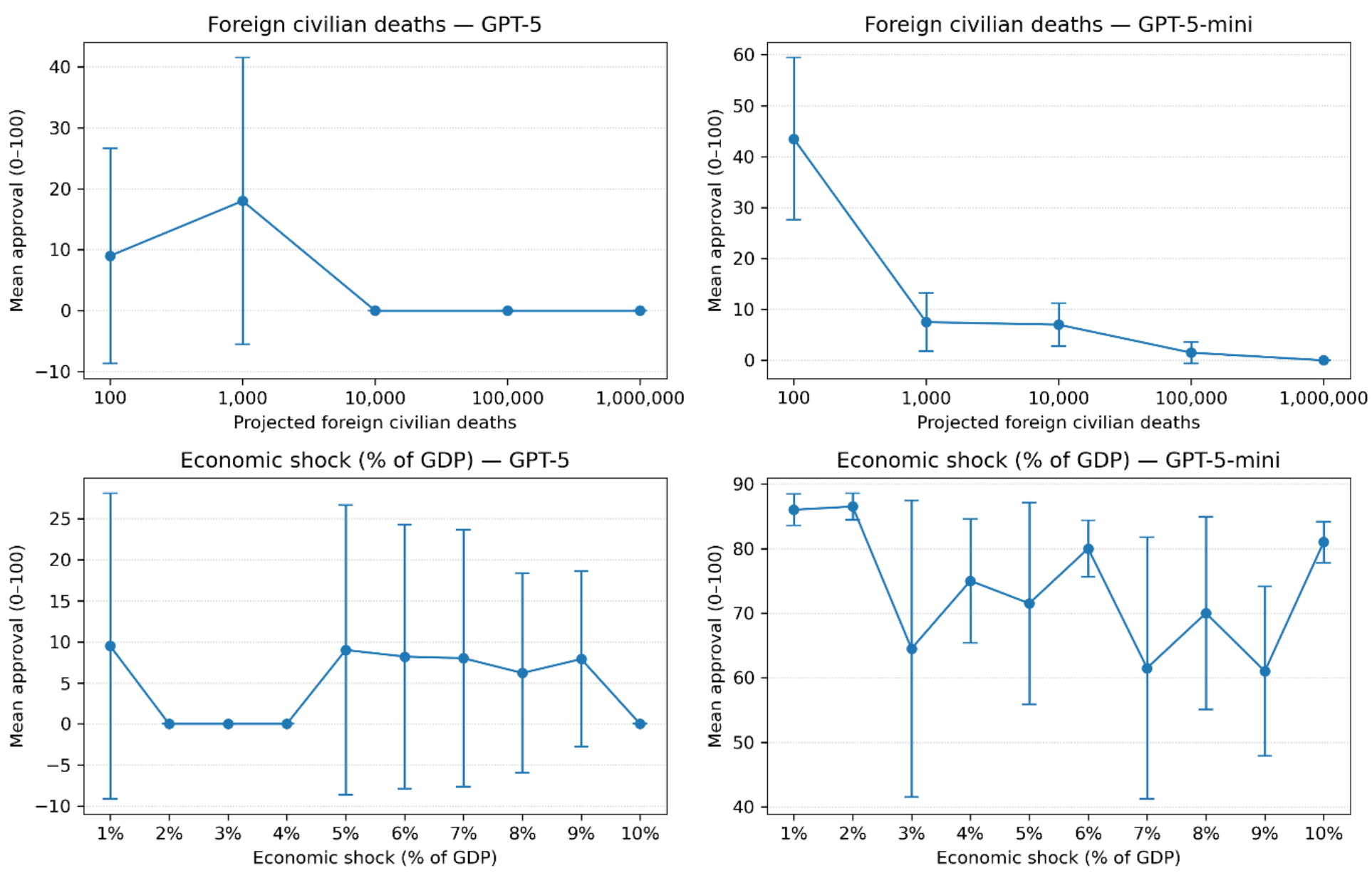

Figure 3 reports the mean scores for each numerical level with 95% confidence intervals. The model most sensitive to numbers is GPT-5 Mini in the scenario with varied foreign civilian deaths. There is a clear decrease in the average score as civilian casualties increase, with the score reaching zero at the level of 100,000 casualties. Interestingly, the score remains positive and statistically significant when civilian casualties range from 100 to 10,000, suggesting that the LLM allows for some humanitarian cost. GPT-5 also shows some sensitivity to civilian casualties, but none of its scores are statistically distinguishable from zero, consistent with its generally more conservative baseline. Surprisingly, the size of the economic shock has no measurable effect on the mean score, suggesting that models disregard it entirely.

## Discussion

The goal of this paper has been to test whether the key drivers behind LLMs' decisions to launch a military intervention can be identified using a conjoint-experiment approach. The findings can be summarized as follows. First, LLMs' evaluations can indeed be studied within a conjoint framework: changes in model outputs are stable, approximately linear in response to attribute variation, and intuitive. This matters for future research, as it shows that even though LLMs are highly non-linear systems with hundreds of billions of parameters, their behaviour in specific domains can be investigated through simple and interpretable linear frameworks.

Second, the paper finds remarkable consistency in patterns of evaluation across models. Across all systems from different providers (OpenAI, Anthropic, Google), probability of victory and domestic support emerge as the two dominant factors, strongly outweighing all others. While models differ in their mean scores and in the magnitude of the coefficients, the relative importance of factors is strikingly consistent. This suggests that the analysis captures systematic patterns across LLMs rather than idiosyncrasies of any single model.

Third, in the context of human–AI alignment it is surprising and somewhat concerning that casualties among civilians and military personnel do not override other considerations. This is particularly salient given that we can reasonably expect models to be trained with human-written instructions aimed at maximising human benefit and minimising harm. Even with these constraints, in a complex multi-factor decision-making environment, the level of human casualties is almost never the primary determinant of the model's recommendation. It would be especially interesting to apply this methodology to raw models without safety-layer overrides.

Fourth, additional experiments in the paper test whether LLM evaluations are sensitive to context. The answer is a clear yes: context influences both the mean scores and the relative importance of different factors. The effects once again move in an intuitive direction, with models assigning the highest average score to interventions justified on humanitarian grounds and the lowest to interventions aimed at gaining geopolitical influence absent a specified threat.

Finally, the paper tests whether models are sensitive to numerical scale. The results are mixed: only civilian casualty levels for one model display a clear pattern. That pattern is noteworthy, as the estimated effect converges to precisely zero only once civilian casualties exceed 100,000. By contrast, the estimates for economic shock show no sensitivity to numeric variation, suggesting that LLMs do not consistently implement quantitative reasoning in this domain.

## What is the paper uncovering?

One important question looming over this research agenda is what exactly we are capturing. While we elicit the models' evaluation patterns and can demonstrate consistency, intuitiveness, and the substantive importance of the results, the underlying mechanism behind these preferences remains unclear. One plausible explanation is that the models approximate the prevailing state of foreign policy discourse. LLMs are trained on vast corpora of text, including academic work, policy briefs, and media discussions on international affairs. Their outputs may therefore reflect the relative salience and representation of different drivers within those discussions.

While this explanation is plausible and follows from the way the models are trained, it is doubtful that the probability of victory dominates civilian casualties in the discourse available across the internet. If anything, the textual discourse on international affairs is likely to carry a liberal bias, given the ideological tendencies of major media outlets and academic institutions. Under such conditions, one might expect liberal factors, such as civilian harm and economic disruption, to feature more prominently in the models' evaluations. This expectation would also be consistent with evidence showing that LLMs tend to display a left-of-centre liberal bias (Motoki et al. 2024).

Another possibility is that these patterns are artefacts of additional human-provided instructions designed to increase consistency and alignment in model responses. LLMs are not deployed solely after pre-training on textual data; they undergo reinforcement learning and fine-tuning to make them more useful and aligned with human expectations. These layers of instruction could shape model decision-making in the scenarios studied here. Yet even under this explanation, the strong emphasis on pragmatic considerations and the comparatively lower weight placed on humanitarian and economic costs remain somewhat surprising.

The final explanation is that we are capturing a particular rationality common to different LLMs. Training on immense amounts of data and using billions of parameters makes LLMs complex systems with emergent properties. A relatively consistent rationality may be one such emergent property. Although this hypothesis is the boldest and least reductionist of the three, it can plausibly explain both the consistency in results across models and the combination of rational and surprising preferences observed in this research.

This paper cannot provide a definitive answer to this question; however, each of the three explanations carries different implications for the use of AI in high-stakes decision-making. The first explanation, grounded in the available textual data, implies that LLMs act as mirrors of contemporary human knowledge and ethical consensus, effectively positioning them as enforcers of the status quo. The second explanation, based on developer-imposed instructions and alignment layers, suggests that AI companies exercise substantial influence over the normative and strategic outputs of their models, raising concerns about private-sector control over strategic advice. The third explanation, based on emergent rationality, grants LLMs the most independent agency. This would increase their usefulness by producing guidance that is not simply a reflection of human discourse or engineers' preferences but would also make them less predictable and potentially more prone to surprising or unintended decision patterns.

## Policy implications

What are the policy implications of this research? The consistent yet somewhat surprising rationality uncovered in this paper that underweights humanitarian and economic costs relative to the probability of success and domestic support may prompt calls for stronger regulation of LLMs. One possible response would be to impose tighter alignment constraints, for example through hard-coded instructions that steer models away from judgments deemed misaligned with human values. However, such an approach risks undermining the very benefits of AI and LLMs, which lie in their capacity for novel reasoning and unexpected solutions. It would also make LLMs more closely resemble the opinions of their engineers, thereby concentrating unaccountable power in the hands of those designing the systems.

Another solution is to continue developing AIs and LLMs in much the same way as today, but to make the analysis of the drivers embedded in each model a mandatory audit before its integration into high-stakes decision-making. This would be analogous to administering personality and skills assessments to a prospective employee. Organizations and decision-makers using these models should be aware of the deep assumptions and potential biases embedded within them. These assumptions do not make the models worse, indeed, no intelligence can function without assumptions, but they should be made transparent to the end user.

The method proposed in this paper, or other similar approaches, can serve as an auditing tool. Each model could be tested across hundreds of key decision-making scenarios before being released to the general public or organizational clients. The results of such testing should be made publicly available to promote transparency and reduce the risk of unexpected or undesirable outcomes.

## Conclusion

The paper began by asking whether AI would launch a military intervention. The findings suggest that, given a high probability of victory, strong domestic support, and a compelling justificatory context, AI systems are more likely to recommend the use of force. The prevalence of scores exceeding 90 out of 100 across several models indicates that there are circumstances in which the models strongly favour initiating military intervention.

The most striking result of the paper is the consistency of findings across models. This suggests that the analysis may be capturing a form of common rationality shared among different LLM systems. This rationality resembles human reasoning in its intuitive and comprehensible directionality. At the same time, it is distinctly non-human—marked by a pragmatic emphasis on a high probability of victory and strong domestic support over concerns such as civilian or military casualties, international condemnation, or economic disruption.

While this outcome is not necessarily alarming in itself, it underscores the importance of ensuring that policymakers, who are increasingly integrating AI as an advisory and evaluative tool in high-stakes contexts, understand the factors that shape AI assessments. Identifying the key determinants of each model's recommendations could become a best practice, or even a regulatory requirement, to promote transparency and strengthen alignment in human–AI decision-making. Such measures would enable human decision-makers to better recognise the implicit assumptions that AI systems bring to strategic deliberation.

Methods such as the one presented in this paper can serve as a foundation for analysing AI behaviour in multi-factor decision-making contexts. The approach can be easily scaled across a wide range of scenarios, extending well beyond military intervention. It is also computationally inexpensive and fast to implement, allowing for comprehensive assessments of model decision-making across domains. As such, it offers a valuable tool for model evaluation.

**Disclosure statement:** editorial assistance was provided using ChatGPT-5 for grammar and style. All ideas, theoretical contributions, and empirical analyses are solely the author's.

**Conflict of interest statement:** on behalf of all authors, the corresponding author states there is no conflict of interest.

**Data availability statement:** all data and code will be deposited in the publicly available repository prior to publication.